\documentstyle[ltwol]{article}
\def\be{\begin{equation}}
\def\ee{\end{equation}}
\def\bea{\begin{eqnarray}}
\def\eea{\end{eqnarray}}
\def\eg{{\em e.g.}}
\def\S{{\mathcal S}}
\def\I{{\mathcal I}}
\def\L{{\mathcal L}}
\def\eff{{\mathrm{eff}}}
\def\Newton{{\mathrm{Newton}}}
\def\bulk{{\mathrm{bulk}}}
\def\matter{{\mathrm{matter}}}
\def\tr{{\mathrm{tr}}}

\def\half{{1\over2}}
\begin{document}

\title{ENERGY CONDITIONS AND THEIR COSMOLOGICAL IMPLICATIONS}

\author{MATT VISSER and CARLOS BARCEL\'O}

\address{Physics Department, Washington University, 
Saint Louis, Missouri 63130-4899, USA
\\
E-mail: visser@kiwi.wustl.edu, carlos@wuphys.wustl.edu
\\
{\em Plenary talk delivered at Cosmo99, Trieste, Sept/Oct 1999}.
\\
{\em gr-qc/0001099; 28 January 2000; \LaTeX-ed \today.}
}


\twocolumn[\maketitle\abstracts{ 
The energy conditions of general relativity permit one to deduce very
powerful and general theorems about the behaviour of strong
gravitational fields and cosmological geometries.  However, the energy
conditions these theorems are based on are beginning to look a lot
less secure than they once seemed: (1) there are subtle quantum
effects that violate all of the energy conditions, and more tellingly
(2), there are also relatively benign looking classical systems that
violate all the energy conditions. This opens up a Pandora's box of
rather disquieting possibilities --- everything from negative
asymptotic mass, to traversable wormholes, to warp drives, up to and
including time machines.
}]

%-------------------------------------------------------------
\section{Introduction}
%-------------------------------------------------------------

Einstein gravity (general relativity) is a tremendously complex theory
even if you restrict attention to the purely classical regime. The
field equations are
\begin{equation}
G^{\mu\nu} = {8\pi \; G_{\mathrm Newton}\over c^4} \; T^{\mu\nu}.
\end{equation}
The left-hand-side, the Einstein tensor $G^{\mu\nu}$, is complicated
enough by itself, but is at least a universal function of the
spacetime geometry. In contrast the right-hand-side, the stress-energy
tensor $T^{\mu\nu}$, is not universal but instead depends on the
particular type of matter and interactions you choose to insert in
your model. Faced with this situation, you must either resign oneself
to performing an immense catalog of special-case calculations, one
special case for each conceivable matter Lagrangian you can write
down, or try to decide on some generic features that ``all
reasonable'' stress-energy tensors should satisfy, and then try to use
these generic features to develop general theorems concerning the
strong-field behaviour of gravitational fields.

One key generic feature that most matter we run across experimentally
seems to share is that energy densities (almost) always seem to be
positive. The so-called ``energy conditions'' of general relativity
are a variety of different ways of making this notion of locally
positive energy density more precise. The (pointwise) energy
conditions take the form of assertions that various linear
combinations of the components of the stress-energy tensor (at any
specified point in spacetime) should be positive, or at least
non-negative. The so-called ``averaged energy conditions'' are
somewhat weaker, they permit localized violations of the energy
conditions, as long as ``on average'' the energy conditions hold when
integrated along null or timelike
geodesics.\cite{Hawking-Ellis,Wald,Book}

The refinement of the energy conditions paralleled the development of
powerful mathematical theorems, such as the singularity theorems
(guaranteeing, under certain circumstances, gravitational collapse
and/or the existence of a big bang
singularity),\cite{Hawking-Ellis,Wald} the positive energy theorem,
the non-existence of traversable wormholes (topological censorship),
and limits on the extent to which light cones can ``tip over'' in
strong gravitational fields (superluminal censorship). All these
theorems require some form of energy condition, some notion of
positivity of the stress-energy tensor as an input hypothesis, and the
variety of energy conditions in use in the relativity community is
driven largely by the technical requirements of how much you have to
assume to easily prove certain theorems.

Over the years, opinions have changed as to how fundamental some of
the specific energy conditions are. One particular energy condition
has now been completely abandoned, and there is general agreement that
another is on the verge of being relegated to the dustbin. There are
however, some more general issues that make one worry about the whole
programme. Specifically:

(1) Over the last decade or so it has become increasingly obvious that
there are subtle quantum effects that are capable of violating {\em
all} the energy conditions, even the weakest of the standard energy
conditions.  Now because these are quantum effects, they are by
definition small (proportional to $\hbar$) so the general consensus
for many years was to not worry too much.\cite{Book}

(2) More recently,\cite{Flanagan-Wald,PLB} it has become clear that
there are quite reasonable looking classical systems, field theories
that are compatible with all known experimental data, and that are in
some sense very natural from a quantum field theory point of view,
which violate all the energy conditions. Because these are now
classical violations of the energy conditions they can be made
arbitrarily large, and seem to lead to rather weird physics. (For
instance, it is possible to demonstrate that Lorentzian-signature
traversable wormholes arise as classical solutions of the field
equations.)\,\cite{PLB}

Faced with this situation, you will either have to learn to live with
some rather peculiar physics, or you will need to make a radical
reassessment of the place of the energy conditions in general
relativity and cosmology.

%\clearpage
%-------------------------------------------------------------
\section{Energy conditions}
%-------------------------------------------------------------

To set some basic nomenclature, the pointwise energy conditions of
general relativity are:\,\cite{Hawking-Ellis,Wald,Book}

\bigskip
\noindent
{\em Trace energy condition} (TEC), {now abandoned}.
\\ 
{\em Strong energy condition} (SEC), {almost abandoned}.
\\ 
{\em Null energy condition} (NEC).
\\ 
{\em Weak energy condition} (WEC).
\\ 
{\em Dominant energy condition} (DEC).
\bigskip

\noindent 
All of these energy conditions can be modified by averaging along null
or timelike geodesics.

The trace energy condition is the assertion that the trace of the
stress-energy tensor should always be negative (or positive depending
on metric conventions), and was popular for a while during the
1960's. However, once it was realized that stiff equations of state,
such as those appropriate for neutron stars, violate the TEC this
energy condition fell into disfavour.\cite{Zeldovich} It has now been
completely abandoned and is no longer cited in the literature --- we 
mention it here as a concrete example of an energy condition being
outright abandoned.

The strong energy condition is currently the subject of much
discussion, sometimes heated. (1) The most naive scalar field theory
you can write down, the minimally coupled scalar field, violates the
SEC,\cite{Hawking-Ellis} and indeed curvature-coupled scalar field
theories also violate the SEC; there are fermionic quantum field
theories where interactions engender SEC violations,\cite{Rose} and
specific models of point-like particles with two-body interactions
that violate the SEC.\cite{Parker} (2) If you believe in cosmological
inflation, the SEC must be violated during the inflationary epoch, and
the need for this SEC violation is why inflationary models are
typically driven by scalar inflaton fields. (3) If you believe the
recent observational data regarding the accelerating universe, then
the SEC is violated on cosmological scales {\em right
now!}~\cite{Science} (4) Even if you are somewhat more conservative,
and regard the alleged present-day acceleration of the cosmological
expansion as ``unproven'', the tension between the age of the oldest
stars and the measured present-day Hubble parameter makes it very
difficult to avoid the conclusion that the SEC must have been violated
in the cosmologically recent past, sometime between redshift 10 and
the present.\cite{Science} Under the circumstances it would be rather
quixotic to take the SEC too seriously as a fundamental guide.

In contrast, the null, weak, and dominant energy conditions are still
extensively used in the general relativity community. The weakest of
these is the NEC, and it is in many cases also the easiest to work
with and analyze. The standard wisdom for many years was that all
reasonable forms of matter should at least satisfy the NEC. After it
became clear that the NEC (and even the ANEC) was violated by quantum
effects two main lines of retrenchment developed:~\cite{Book} 

(1) Many researchers simply decided to ignore quantum mechanics,
relying on the classical NEC to prevent grossly weird physics in the
classical regime, and hoping that the long sought for quantum theory
of gravity would eventually deal with the quantum problems. This is
not a fully satisfactory response in that NEC violations already show
up in semiclassical quantum gravity (where you quantize the matter
fields and keep gravity classical), and show up at first order in
$\hbar$. Since semiclassical quantum gravity is certainly a good
approximation in our immediate neighborhood, it is somewhat disturbing
to see widespread (albeit small) violations of the energy conditions
in the here and now. Many experimental physicists and observational
astrophysicists react quite heatedly when the theoreticians tell them
that according to our best calculations there should be plenty of
``negative energy'' (energy densities less than that of the flat-space
Minkowski vacuum) out there in the real universe. However, to avoid
the conclusion that quantum effects can and do lead to locally
negative energy densities, and even violations of the ANEC, requires
truly radical surgery to modern physics, and in particular we would
have to throw away almost all of quantum field theory.

(2) A more nuanced response is based on the Ford--Roman {\em Quantum
Inequalities}.\cite{Ford-Roman} These inequalities, which are
currently still being developed and extended, and whose implications
are still a topic of considerable activity, are based on the facts
that while quantum-induced violations of the energy conditions are
widespread they are also {\em small}, and on the observation that a
negative energy in one place and time always seems to be compensated
for (indeed, over-compensated for) by positive energy elsewhere in
spacetime. This is the so-called {\em Quantum Interest
Conjecture}.\cite{Ford-Roman} While the positive pay-back is not
enough to prevent violation of the ANEC (based on averaging the NEC
along a null geodesic) the hope is that it will be possible to prove
some type of space-time averaged energy condition from first
principles, and that such a space-time averaged energy condition might
be sufficient to enable us to recover the
singularity/positive-mass/censorship theorems under weaker hypotheses
than currently employed.

A fundamental problem for this type of approach is the more recent
realization that there are also serious {\em classical} violations of
the energy conditions.\cite{Flanagan-Wald,PLB} These classical
violations can easily be made arbitrarily large, and appear to be
unconstrained by any form of energy inequality. The simplest source of
classical energy condition violations is from scalar fields, so we
shall first present some background on the usefulness and need for
scalar field theories in modern physics.

%\clearpage

%-------------------------------------------------------------
\section{Scalar Fields: Background}
%-------------------------------------------------------------

Scalar fields play a somewhat ambiguous role in modern theoretical
physics: on the one hand they provide great toy models, and are from a
theoretician's perspective almost inevitable components of any
reasonable model of empirical reality; on the other hand the direct
experimental/observational evidence is spotty.

The only scalar fields for which we have really direct ``hands-on''
experimental evidence are the scalar mesons (pions $\pi$; kaons $K$;
and their ``charmed'', ``truth'' and ``beauty'' relatives, plus a
whole slew of resonances such as the $\eta$, $f_0$, $\eta'$, $a_0$,
\dots).\cite{PDG} 
Not one of these particles are fundamental, they are all
quark-antiquark bound states, and while the description in terms of
scalar fields is useful when these systems are probed at low momenta
(as measured in their rest frame) we should certainly not continue to
use the scalar field description once the system is probed with
momenta greater than $\hbar/({\mathrm{bound~state~radius}})$. In terms
of the scalar field itself, this means you should not trust the scalar
field description if gradients become large, if
\begin{equation}
||\nabla \phi|| > {||\phi||\over {\mathrm{bound~state~radius}} }.
\end{equation}
Similarly you should not trust the scalar field description if the
energy density in the scalar field exceeds the critical density for
the quark-hadron phase transition. Thus scalar mesons are a mixed bag:
they definitely exist, and we know quite a bit about their properties,
but there are stringent limitations on how far we should trust the
scalar field description.

The next candidate scalar field that is closest to experimental
verification is the Higgs particle responsible for electroweak
symmetry breaking.  While in the standard model the Higgs is
fundamental, and while almost everyone is firmly convinced that some
Higgs-like scalar field exits, there is a possibility that the
physical Higgs (like the scalar mesons) might itself be a bound state
of some deeper level of elementary particles (\eg, technicolor and its
variants). Despite the tremendous successes of the standard model of
particle physics we do not (currently) have direct proof of the
existence of a fundamental Higgs scalar field.

A third candidate scalar field of great phenomenological interest is
the axion: it is extremely difficult to see how one could make strong
interaction physics compatible with the observed lack of strong CP
violation, without something like an axion to solve the so-called
``strong CP problem''. Still, the axion has not yet been directly
observed experimentally.

A fourth candidate scalar field of phenomenological interest
specifically within the astrophysics/cosmology community is the
so-called ``inflaton''. This scalar field is used as a mechanism for
driving the anomalously fast expansion of the universe during the
inflationary era. While observationally it is a relatively secure bet
that something like cosmological inflation (in the sense of
anomalously fast cosmological expansion) actually took place, and
while scalar fields of some type are the only known reasonable way of
driving inflation, we must again admit that direct observational
verification of the existence of the inflaton field (and its variants,
such as quintessence) is far from being accomplished.

A fifth candidate scalar field of phenomenological interest
specifically within the general relativity community is the so-called
``Brans--Dicke scalar''. This is perhaps the simplest extension to
Einstein gravity that is not ruled out by experiment. (It is certainly
greatly constrained by observation and experiment, and there is no
positive experimental data guaranteeing its existence, but it is not
ruled out.) The relativity community views the Brans--Dicke scalar
mainly as an excellent testing ground for alternative ideas and as a
useful way of parameterizing possible deviations from Einstein
gravity. (And experimentally and observationally, Einstein gravity
still wins.)

Finally, the membrane-inspired field theories (low-energy limits of
what used to be called string theory) are literally infested with
scalar fields. In membrane theories it is impossible to avoid scalar
fields, with the most ubiquitous being the so-called
``dilaton''. However, the dilaton field is far from unique, in
general there is a large class of so-called ``moduli'' fields, which
are scalar fields corresponding to the directions in which the
background spacetime geometry is particularly ``soft'' and easily
deformed. So if membrane theory really is the fundamental theory of
quantum gravity, then the existence of fundamental scalar fields is
automatic, with the field theory description of these fundamental
scalars being valid at least up to the Planck scale, and possibly
higher. 

(For good measure, by making a conformal transformation of the
spacetime geometry it is typically possible to put membrane-inspired
scalar fields into a framework which closely parallels that of the
generalized Brans--Dicke fields. Thus there is a potential for much
cross-pollination between Brans--Dicke inspired variants of general
relativity and membrane-inspired field theories.)

So overall, we have excellent theoretical reasons to expect that
scalar field theories are an integral part of reality, but the direct
experimental/observational verification of the existence of
fundamental fields is still an open question. Nevertheless, we think
it fair to say that there are excellent reasons for taking scalar
fields seriously, and excellent reasons for thinking that the
gravitational properties of scalar fields are of interest
cosmologically, astrophysically, and for providing fundamental probes
of general relativity.

%-------------------------------------------------------------
\section{Scalar Fields and Gravity}
%-------------------------------------------------------------

In setting up the formalism for a scalar field coupled to gravity we
first need to specify a few conventions: the metric signature will be
taken to be $(-,+,+,+)$ and we adopt Landau--Lifshitz spacelike
conventions (LLSC) This is equivalent to MTW~\cite{MTW} conventions
with latin indices for the tensors, and is equivalent to
Flanagan--Wald.\cite{Flanagan-Wald} In the MTW classification this
corresponds to $(+g,+Riemann,+Einstein)$. We take the total action to
be:
\begin{equation}
\S = \S_g + \S_\phi + \S_\bulk.
\end{equation}
The gravity action is standard
\begin{equation}
\S_g = \int d^4 x \sqrt{-g} \; \half \; \kappa \; R,
\end{equation}
with the ordinary Newton constant being defined by
\begin{equation}
\kappa = {c^4 \over 8\pi\; G_\Newton}.
\end{equation}
We shall permit the scalar field to exhibit an arbitrary curvature coupling $\xi$
\begin{equation}
\S_\phi =   \int d^d x \sqrt{-g} \;
\left(
- \half \left[ (\nabla\phi)^2 + \xi R \phi^2 \right]
- V(\phi)
\right).
\end{equation}
Finally the action for ordinary bulk matter is taken as
\begin{equation}
\S_\bulk =   \int d^d x \sqrt{-g} \; f(\phi) \; \L_\matter.
\end{equation}
Here we assume $f(\phi)$ is algebraic, and that $\L_\matter$ does not
involve $\phi$. We also assume for technical reasons that $\L_\matter$
does not contain any terms involving second derivatives of the
metric. (For example, let $L_\matter$ be the Lagrangian of the
ordinary standard model of particle physics.)  Under these
circumstances the equations of motion (EOM) for gravity can be written
\begin{equation}
\kappa \; G_{ab} = T^\phi_{ab} + f(\phi) \; T^\matter_{ab}.
\end{equation}
The EOM of the $\phi$ field are
\begin{equation}
(\nabla^2 - \xi R) \phi - V'(\phi)+ f'(\phi) \; \L_\matter=0.
\end{equation}
The EOM for the bulk matter fields can be phrased as
\begin{equation}
\nabla_b \left[ f(\phi) \; T_\matter^{ab} \right] = 0.
\end{equation}
There are a few hidden subtleties here: First because of the curvature
coupling term $\xi R \phi^2$ the stress-energy tensor for the scalar
field contains a term proportional to the Einstein tensor
\begin{eqnarray}
T_{ab}^\phi &=& 
\nabla_a \phi \; \nabla_b \phi 
- \half g_{ab} (\nabla\phi)^2 
- g_{ab} V(\phi)
\\
&+& 
\xi\left[
G_{ab} \; \phi^2 - 2\nabla_a(\phi \; \nabla_b\phi) + 
2 g_{ab} \nabla^c(\phi \nabla_c \phi)
\right].
\nonumber
\end{eqnarray}
Second, the way we have defined $T^\matter$ it is to be calculated
from $\S_\bulk$ by simply ignoring the factor $f(\phi)$.

Because the RHS contains a term proportions to the Einstein tensor it
is best to rearrange the gravity EOM by isolating all such terms on
the LHS, in which case the gravity EOM is equivalent to
\begin{equation}
\kappa \; G_{ab} = [T^\phi_\eff]_{ab} + 
f(\phi)\; {\kappa \over\kappa-\xi\phi^2}\; T^\matter_{ab}.
\end{equation}
Here the ``effective'' stress-energy for the scalar field is
\begin{eqnarray}
[T^\phi_\eff]_{ab} &=& {\kappa\over\kappa-\xi\phi^2} 
\Bigg\{
\nabla_a \phi \; \nabla_b \phi - \half g_{ab} (\nabla\phi)^2 
- g_{ab} V(\phi)
\nonumber\\
&-&
\xi\left[
2\nabla_a(\phi \; \nabla_b\phi) -
2 g_{ab} \nabla^c(\phi \nabla_c \phi)
\right] \Bigg\}.
\end{eqnarray}
Note that this effective stress-energy can change sign for certain
values of the scalar field, which is our first hint of peculiar
physics. More importantly, observe that it is the energy conditions
defined in terms of this effective stress-energy tensor that are the
physically interesting ones: energy conditions imposed upon this
effective stress-energy imply constraints on the Einstein tensor,
placing constraints on the curvature, which is what is really needed
for deriving singularity/positive-mass/censorship theorems.

In addition, the ``effective'' gravitational coupling of ``normal
matter'' to the gravitational field is
\begin{equation}
\kappa_\eff = {\kappa-\xi\phi^2\over f(\phi)}.
\end{equation}
In terms of Newton's constant
\begin{equation}
G^\eff_\Newton(\phi) = G_\Newton  \; f(\phi)\; {\kappa \over\kappa-\xi\phi^2}.
\end{equation}
It is particularly convenient to pick $f(\phi) =
(\kappa-\xi\phi^2)/\kappa$.  With this simplifying choice bulk matter
couples to gravity in the ordinary way, while the only gravitational
peculiarities are now concentrated in the gravity-$\phi$ sector. 

In the remainder of the technical discussion we shall, for simplicity,
treat the normal matter in the test-field limit. That is, whatever
normal matter is present is assumed to be sufficiently diffuse to not
appreciably affect the spacetime geometry, while test particles of
normal matter can still be used as convenient probes of the spacetime
geometry.  Making this test-matter approximation thus simplifies life
to the extent that we are looking at a gravity EOM
\begin{equation}
\kappa \; G_{ab} = [T^\phi_\eff]_{ab},
\end{equation}
coupled to the scalar EOM
\begin{equation}
(\nabla^2 - \xi R) \phi - V'(\phi)=0.
\end{equation}
This system is now sufficiently simple that a number of exact analytic
solutions are known.\cite{PLB,Fisher,Bergmann,Ellis} We shall not
re-derive any of these exact analytic solutions but instead will quote
them as examples when we want to illustrate some aspect of the generic
situation.\cite{PLB}

%-------------------------------------------------------------
\section{Scalar Fields and energy conditions}
%-------------------------------------------------------------

%-------------------------------------------------------------
\subsection{SEC}
%-------------------------------------------------------------

Technically the SEC is defined (in 4 dimensions) by using the
trace-reversed stress energy tensor
\begin{equation}
\overline{T}_{ab} \equiv T_{ab} - \half \; g_{ab} \; \tr(T).
\end{equation}
The SEC is then the assertion that for any timelike vector
\begin{equation}
\overline{T}_{ab} \; v^a \;  v^b \geq 0.
\end{equation}
The reason for wanting this condition is that, via the Einstein
equations, the SEC would imply the Ricci convergence condition
\begin{equation}
R_{ab} \; v^a \;  \geq 0.
\end{equation}
This convergence condition on the Ricci curvature tensor is then used
to prove that nearby timelike geodesics are always focussed towards
each other, and this focussing lemma is then a first critical step in
proving singularity theorems and the like. Unfortunately if you
actually calculate this quantity for the effective stress-energy of
the scalar field you find
\begin{eqnarray}
\overline{[T^\phi_\eff]}_{ab} \; v^a \; v^b &=&
{\kappa\over\kappa-\xi\phi^2}
\Bigg\{
(v^a \nabla_a \phi)^2 
- V(\phi)
\\ 
&-&
\xi\left[
2 v^a v^b\nabla_a(\phi \nabla_b\phi) -  \nabla^c(\phi \nabla_c \phi)
\right]
\Bigg\}.
\nonumber
\end{eqnarray}
It's very easy to make this quantity negative. There is a specific
example in Hawking--Ellis,\cite{Hawking-Ellis} page 95; flat space
with $\xi=0$ and $V(\phi) = \half m^2 \phi^2$, but the phenomenon is
much more general.  For minimal coupling ($\xi=0$) we see
\begin{equation}
\overline{[T^\phi_\eff]}_{ab} \; v^a \; v^b =
(v^a \nabla_a\phi)^2 - V(\phi).
\end{equation}
Thus any static field with a positive potential violates the SEC. (For
example, a slowly changing scalar with a mass or quartic
self-interactions, this is the simplest toy model for cosmological
inflation and/or quintessence; also any {\em positive} cosmological
constant violates the SEC.)

Adding a non-minimal coupling ($\xi\neq0$) is not helpful, though it
does make the algebra messier. (In particular, adding non-minimal
coupling will not by itself switch off cosmological inflation.)  The
key point regarding the SEC is that a positive potential energy
$V(\phi)>0$ tends to violate the SEC, and you can always think of
getting SEC violations by going to a region of field space with high
potential energy.

Though the SEC violations we are talking about here are most commonly
used in building models of cosmological inflation, perhaps the most
serious long term issue is that the singularity theorem we use to
prove the existence of a big bang singularity in FLRW cosmologies
depends crucially on the SEC. Not only do SEC violations permit
cosmological inflation but they also open the door to replacing the
big bang singularity with a ``bounce'', or ``Tolman
wormhole''.\cite{Rose,Parker,Tolman} In fact it is now known that
generically the SEC is the {\em only} energy condition you need to
violate to get a bounce, and that it is still possible to satisfy all
the other energy conditions at and near the bounce.\cite{Tolman} (In
counterpoint, if one replaces the SEC by strong enough inhomogeneity
assumptions then it is possible to prove that certain classes of
chaotic inflationary models must nevertheless possess a big-bang
singularity.)

Note that scalar fields do not guarantee the prevention of a big bang
singularity, they merely raise this possibility --- and this issue is
interesting enough to warrant further investigation. Another
interesting point is that the reason that most inflationary models are
based on minimally coupled scalars ($\xi=0$) is purely a historical
one, there simply was no need to go to non-minimal coupling to get the
SEC violations that are needed for cosmological inflation, and it is
only once you get down to rather specific model building that
non-minimal coupling becomes interesting.

%-------------------------------------------------------------
\subsection{NEC}
%-------------------------------------------------------------

The NEC has two great advantages over the SEC: it is the simplest
energy condition to deal with algebraically, and because it is the
weakest pointwise energy condition it leads to the strongest
theorems. The NEC is the assertion that for any null vector $k^a$ we
should have
\begin{equation}
{T}_{ab} \; k^a \;  k^b \geq 0.
\end{equation}
Unfortunately when we actually calculate this quantity for the scalar
field we find
\begin{eqnarray}
[T^\phi_\eff]_{ab} \; k^a \; k^b &=&
{\kappa\over\kappa-\xi\phi^2} 
\Bigg\{
(k^a \nabla_a \phi)^2 
\nonumber\\
&& 
- \xi\left[
2 k^a k^b \nabla_a (\phi \nabla_b\phi)
\right]
\Bigg\}.
\end{eqnarray}
For minimal coupling
\begin{equation}
[T^\phi_\eff]_{ab} \; k^a \; k^b =
(k^a \nabla_a \phi)^2 \geq 0.
\end{equation}
The accident that the NEC is satisfied for minimal coupling led to a
situation where researchers just did not look under enough rocks to see
where the problems lay.

For non-minimal coupling we start, as a convenience, by extending $k$
to be a geodesic vector field around the point of interest, so that
$k^a \nabla_a k^b = 0$. Then using the affine parameter $\lambda$ we
have $k^a \nabla_a = d/d\lambda$ so that
\begin{equation}
[T^\phi_\eff]_{ab} \; k^a \; k^b =
{\kappa\over\kappa-\xi\phi^2} 
\left\{
\left({d\phi\over d\lambda}\right)^2 -
\xi \left({d^2[\phi^2]\over d\lambda^2}\right)
\right\}.
\end{equation}
Pick any local extremum of $\phi$ along the null geodesic, then
\begin{equation}
[T^\phi_\eff]_{ab} \; k^a \; k^b =
- {\kappa\over\kappa-\xi\phi^2} 
\left\{
\xi  \left({d^2[\phi^2]\over d\lambda^2}\right)
\right\}.
\end{equation}
It is easy to make this negative (for \emph{any} $\xi\neq0$).  If
$\xi<0$ consider any local maximum of the field $\phi$, the NEC is
violated. If $\xi>0$ and $\phi<\sqrt{\kappa/\xi}$ then any local
minimum does the job, while if $\xi>0$ and $\phi>\sqrt{\kappa/\xi}$
one again needs a local maximum of $\phi$ to violate the NEC.

Now classical violations of the NEC are much more disturbing than
classical violations if the SEC. In particular, traversable
Lorentzian-signature wormholes are known to be associated with
violations of the NEC (and ANEC),\cite{Morris-Thorne} as are
warp-drives,\cite{Bondi,Alcubierre} time machines,\cite{Book} and
similar exotica --- this should start to make you feel just a little
nervous.

%-------------------------------------------------------------
\subsection{ANEC}
%-------------------------------------------------------------

The ANEC is technically much more interesting. Pick some complete null
geodesic $\gamma$ and consider the integral
\begin{equation}
\I = 
\oint
[T^\phi_\eff]_{ab} \; k^a \; k^b \; d\lambda.
\end{equation}
Then
\begin{equation}
\I = 
\oint
{\kappa\over\kappa-\xi\phi^2} 
\left\{
\left({d\phi \over d\lambda}\right)^2 -
2 \xi
{d\over d\lambda} \left( \phi {d\phi\over d\lambda} \right)
\right\} d\lambda.
\end{equation}
Integrate by parts, discarding the boundary terms (that is, assume
sufficiently smooth asymptotic behaviour)
\begin{equation}
\I = 
\oint
{\kappa\over\kappa-\xi\phi^2} 
\left\{
\left({d\phi \over d\lambda}\right)^2  +
{
4\xi^2 \phi^2  (d\phi/d\lambda)^2
\over
\kappa-\xi\phi^2
}
\right\} d\lambda.
\end{equation}
Now assemble the pieces:
\begin{equation}
\I = 
\oint
{\kappa [\kappa - \xi(1-4\xi)\phi^2 ]\over(\kappa-\xi\phi^2)^2} 
\left({d\phi \over d\lambda}\right)^2 
d\lambda.
\end{equation}
The integrand is {\em not} positive definite, and ANEC can be
violated, provided there is a region along the geodesic where
\begin{equation}
\xi(1-4\xi)\; \phi^2 > \kappa. 
\end{equation}
This can only happen for $\xi \in (0, 1/4)$, so there is something
very special about this range of curvature couplings. In particular
$1/6 \in (0,1/4)$, so conformal coupling in $d=4$ lies in this range.
This is important because there are technical issues in quantum field
theory that seem to almost automatically imply that real physical
scalars should be conformally coupled. (Conformal coupling $\xi=1/6$
is typically a infrared fixed point of the renormalization group flow,
furthermore setting $\xi=1/6$ and then going to flat spacetime
automatically reproduces the so-called ``new-improved stress-energy
tensor'', a stress-energy tensor that is much better behaved in
quantum field theory than the unimproved minimally-coupled
stress-energy tensor.)\,\cite{PLB}

Also note that ANEC violations require
\begin{equation}
\phi^2 > {\kappa\over\xi(1-4\xi)} > {\kappa\over\xi}.
\end{equation}
This implies that the prefactor in the effective stress-energy tensor
for scalars must go through a zero and become negative in order to get
ANEC violations. Thus ANEC violations are considerably more
constrained than NEC violations, and the ANEC violating regions are
always associated with regions where the effective stress-energy has
``reversed sign''.

Furthermore in the ANEC violating region 
\begin{equation}
\phi^2 > {\kappa\over\xi(1-4\xi)} > 16 \kappa, 
\end{equation}
implying that the scalar field must take on enormous
(super--Planckian) values in order to provide ANEC violations. Now
super--Planckian values for scalar fields are not that unusual, they
are part and parcel of many (though not all) inflationary models, and
the standard lore in cosmology is to not worry about a super-Planckian
{\em value} for the scalar field unless the energy density is also
super--Planckian.

%-------------------------------------------------------------
\section{Conclusions}
%-------------------------------------------------------------

Even with the caveats provided above, the fact that the ANEC can be
violated by classical scalar fields is significant and important ---
in particular the ANEC is the weakest of the energy conditions in
current use, and violating the ANEC short circuits {\em all} the
standard singularity/positive-mass/censorship theorems. This
observation piqued our interest and we decided to see just how weird
the physics could get once you admit scalar fields into your
models.\cite{PLB}

In particular, it is by now well-known that traversable wormholes are
associated with violations of the NEC and ANEC, so we became
suspicious that there might be an explicit class of exact traversable
wormhole solutions to the coupled gravity-scalar field
system.\cite{Book} In a recent paper\,\cite{PLB} we presented such a
class of solutions --- for algebraic simplicity we restricted
attention to conformal coupling ($\xi \to 1/6$) and set the potential
to zero $V(\phi)\to 0$. We found a three-parameter class of exact
solutions to the coupled Einstein--scalar field equations with the
three parameters being the mass, the scalar charge, and the value of
the scalar field at infinity. Within this three-dimensional parameter
space we found a two-dimensional subspace that corresponds to exact
traversable wormhole solutions.\cite{PLB}

The simplifications of conformal coupling and zero potential were made
only for the sake of algebraic simplicity and we expect that there are
more general classes of wormhole solutions waiting to be
discovered. In addition, deviations from spherical symmetry are also
of interest. It should be borne in mind that although the present
analysis indicates that there must be super-Planckian scalar fields
{\em somewhere} in the wormhole spacetime, this does not necessarily
mean that a traveller needs to traverse one of these super-Planckian
regions to cross to the other side: If the wormhole is not spherically
symmetric it is typically possible to minimize the spatial extent of
the regions of peculiar physics.\cite{Simple} Work on these topics is
continuing.

Now traversable wormholes, while certainly exotic, are by themselves
not enough to get the physics community really upset: The big problem
with traversable wormholes is that if you manage to acquire even one
inter-universe traversable wormhole then it {\em seems} almost
absurdly easy to build a time machine --- and this does get the
physics community upset.\cite{Book} There is a conjecture ({\em
Hawking's Chronology Protection Conjecture}) that quantum physics will
save the universe by destabilizing the wormhole just as the time
machine is about to form, but it must certainly be emphasized that
there is considerable uncertainty as to how serious these causality
problems are.\cite{Book}

There are two responses to the current state of affairs: either we can
learn to live with wormholes, and other strange physics engendered by
energy condition violations, or we need to patch up the theory. We
cannot just say that some improved version of the energy conditions
will do the job for us, since we already have an explicit solution of
the Einstein equations that contains traversable wormholes --- we
would have to do something more drastic, like attack the notion of a
scalar field, or forbid conformal couplings [we would need to forbid
the entire range $\xi\in(0,1/4)$], or forbid super-Planckian field
values --- each one of these particular possibilities however is in
conflict with cherished notions of some segments of the particle
physics/ membrane theory/ relativity/ astrophysics communities. Most
physicists would be loathe to give up the notion of a scalar field,
and conformal coupling is so natural that it is difficult to believe
that banning it would be a viable option. Banishing super--Planckian
field values is more plausible, but this runs afoul of at least some
segments of the inflationary community.

%-------------------------------------------------------------
\section*{References}
%-------------------------------------------------------------

%-------------------------------------------------------------

\begin{thebibliography}{99}
%-------------------------------------------------------------
\bibitem{Hawking-Ellis}
S.W. Hawking and G.F.R. Ellis, 
{\em The large scale structure of spacetime},
(Cambridge University Press, England, 1973).

\bibitem{Wald}
R.M. Wald, 
{\em General Relativity},
(University of Chicago Press, Chicago, 1984). 

\bibitem{Book} 
M. Visser, 
{\em Lorentzian wormholes}, 
(AIP Press, New York, 1995).

\bibitem{Flanagan-Wald}
E.E. Flanagan and R.M. Wald, Phys. Rev. D {\bf 54}, 6233 (1996). 

\bibitem{PLB}
C. Barcel\'o and M. Visser, Phys. Lett. {\bf B466}, 127 (1999).

\bibitem{Zeldovich}
Ya.B. Zeldovich, and I.D. Novikov,
{\em Stars and Relativity (Relativistic Astrophysics, Vol. 1),}
(University of Chicago Press, Chicago, 1971), see esp. p. 197.

\bibitem{Rose}
B. Rose, Class. Quantum Grav. {\bf 3}, 975 (1986); {\bf 4}, 1019 (1987). 

\bibitem{Parker}
L. Parker and Y. Wang, Phys. Rev. D {\bf 42}, 1877 (1990).

\bibitem{Science}
M. Visser,
Science {\bf 276}, 88 (1997).

\bibitem{Ford-Roman}
L.H. Ford and T.A. Roman,
Phys. Rev. D {\bf51}, 4277 (1995); D {\bf 60}, 104018 (1999); \\
L.H. Ford, M.J. Pfenning, and T.A. Roman,
Phys. Rev. D {\bf57}, 4839 (1998). 

\bibitem{PDG}
Particle Data Group, {\sf http://pdg.lbl.gov}, 
Euro. Phys. J. {\bf C3}, 1 (1998).

\bibitem{MTW}
C.W. Misner, K.S. Thorne, and J.A. Wheeler, {\em Gravitation},
(Freeman, San Francisco, 1973).

\bibitem{Fisher}
I.Z. Fisher, Zh. Exp. Th. Fiz. {\bf 18}, 636 (1948); gr-qc/9911008.

\bibitem{Bergmann}
O. Bergmann and R. Leipnik, Phys. Rev. {\bf 107}, 1157 (1957).

\bibitem{Ellis}
H. Ellis, J. Math. Phys. {\bf 14}, 104 (1973); 
Errata {\bf 15}, 520 (1974).

\bibitem{Tolman}
D. Hochberg, C. Molina--Par\'\i{}s, and M. Visser, 
Phys. Rev. D {\bf 59}, 044011 (1999); \\
C. Molina--Par\'\i{}s and and M. Visser, 
Phys. Lett. {\bf B455}, 90 (1999).


\bibitem{Morris-Thorne} 
M.S. Morris and K.S. Thorne, Am. J. Phys. {\bf 56}, 395 (1988).


\bibitem{Bondi}
H. Bondi, Rev. Mod. Phys. {\bf 29}, 423 (1957).

\bibitem{Alcubierre}
M. Alcubierre, Class. Quantum Grav. {\bf 11}, L73 (1994).

\bibitem{Simple}
M. Visser, Phys. Rev. D {\bf 39}, 3182 (1989).
 

%-------------------------------------------------------------
\end{thebibliography}
\end{document}